\documentclass[11pt,a4paper]{article}

\usepackage{bm,amsfonts,amssymb,mathrsfs,amsmath}
\usepackage{graphicx}
\usepackage[colorlinks=true]{hyperref}
\usepackage{ulem}
\usepackage{color}
\usepackage{comment}

\newcommand{\U}{\mathrm{U}}

\newcommand{\SU}{\mathrm{SU}}

\newcommand{\rem}[1]{\textcolor{blue}{\bf[#1]}}

\newcommand{\beq}{\begin{eqnarray}}
\newcommand{\eeq}{\end{eqnarray}}

\newcommand{\bea}{\begin{eqnarray}}
\newcommand{\eea}{\end{eqnarray}}

\makeatletter
\@addtoreset{equation}{section}
\makeatother

\date{\today}

\begin{document}

\begin{titlepage}

\renewcommand{\thefootnote}{\fnsymbol{footnote}}

\begin{flushright}
\end{flushright}

\vskip9em

 \hspace{-2.5em}
 {\LARGE\textbf{Conformal field theory analysis for QCD Kondo effect}}
 
\begin{center}

 \vskip5em

 \setcounter{footnote}{1}
 {\sc Taro Kimura}$^1$\footnote{E-mail address: 
 \href{mailto:taro.kimura@keio.jp}
 {\tt taro.kimura(at)keio.jp}}
 and
 {\sc Sho Ozaki}$^{1,2}$\footnote{E-mail address: 
 \href{mailto:shoozaki0117@gmail.com}
 {\tt shoozaki0117(at)gmail.com}}

 \vskip2em

\textit{
 1. Department of Physics \& Research and Education Center for Natural Sciences, \\
 Keio University, Kanagawa
 223-8521, Japan\\
 2. Department of Radiology, University of Tokyo Hospital, Tokyo 113-8655, Japan
}

 \vskip3em

 \end{center}

 \vskip2em

 \begin{abstract}
  We study non-perturbative aspects of QCD Kondo effect, which has been recently proposed for the finite density and strong magnetic field systems, using conformal field theory describing the low energy physics near the IR fixed point.
  We clarify the symmetry class of QCD Kondo effect both for the finite density and magnetic field systems, and show how the IR fixed point is non-perturbatively characterized by the boundary condition, which incorporates the impurity effect in Kondo problem.
  We also obtain the low temperature behavior of several quantities of QCD Kondo effect in the vicinity of the IR fixed point based on the conformal field theory analysis.
 \end{abstract}

\end{titlepage}

\tableofcontents

\hrulefill

\vspace{1em}

\setcounter{footnote}{0}

\section{Introduction}
\label{sec:intro}

Recently, a novel type of the Kondo effect induced by color degrees of freedom, so-called QCD Kondo effect, is proposed \cite{Hattori:2015hka}.
QCD Kondo effect is a Kondo effect realized in high density quark matter with a heavy quark impurity.
It is well known that there are three important ingredients for the appearance of the Kondo effect: (i) Fermi surface, (ii) quantum fluctuations (loop effects), (iii) non-Abelian property of interaction. 
In the QCD Kondo effect, the last condition (iii) corresponds to the color exchange interaction mediated by gluon between a light quark near the Fermi surface and the heavy quark impurity.
Near the Fermi surface, the system becomes effectively (1+1)-dimensional.
This dimensional reduction plays an essential role for the appearance of the Kondo effect.
As a later development of the QCD Kondo effect, one of the authors together with the others have proposed the magnetically induced QCD Kondo effect \cite{Ozaki:2015sya}.
In strong magnetic field, the dimensional reduction to (1+1)-dimensions also occurs.
This (1+1)-dimensional dynamics gives rise to magnetically induced QCD Kondo effect.

A lot of approaches to the non-perturbative regime of the Kondo effect have been developed, since the standard perturbative analysis does not work below the typical energy scale, due to the asymptotic freedom.
The Conformal Field Theory (CFT) is one of such approaches to study the IR fixed point of the Kondo effect~\cite{Affleck:1990zd,Affleck:1991tk,Affleck:1990by,Affleck:1990iv,Affleck:1992ng,Ludwig:1994nf}.
See also a review article~\cite{Affleck:1995ge}.
In the CFT approach, the impurity effect is treated as the boundary condition. 
Thus we can discuss the non-trivial boundary behavior using the boundary CFT method.
This approach allows us to compute the boundary contribution to the entropy, which characterizes the ground state degeneracy of the impurity, called the $g$-factor.
This $g$-factor gives a useful information about the IR fixed point of the Kondo problem.
If this degeneracy is given by an integer value, the system would be described by the Fermi liquid.
On the other hand, when the degeneracy becomes irrational, which is typically observed in the over-screening system, it could be a signal of the non-Fermi liquid nature.
In addition to the $g$-factor, we can obtain the low temperature behavior of several quantities, e.g., specific heat, susceptibility.
Usually, to obtain such scaling behaviors, we have to compute the correlation functions and their scaling limit.
Thus it is difficult to evaluate it in general.
However, in CFT approach this scaling behavior can be studied using the perturbation analysis with respect to the leading irrelevant operator in the vicinity of the IR fixed point.
Similarly we can also compute the exact value of the Wilson ratio, which is known as the universal quantity of the Kondo system at the IR fixed point.

In this paper we apply the CFT approach to QCD Kondo problem both for the finite density and strong magnetic field systems.
We first clarify the effective (1+1)-dimensional field theory describing the low-energy physics, especially with emphasis on some specific features of the QCD Kondo effect. 
We will show that the symmetry at the IR fixed point is enhanced from that for the UV theory, which is indeed a specific feature of the QCD Kondo effect, since such a symmetry enhancement does not occur in the conventional Kondo effect.
We then discuss the $g$-factor and its dependence on the flavor degrees of freedom.
It shows that the IR fixed point would exhibit the non-Fermi liquid behavior for the finite density system, while the Fermi liquid for the strong magnetic field system.
We also discuss the low temperature behavior of some quantities just by applying the established formula.

The remaining part of this article is organized as follows.
In Sec.~\ref{sec:CFT_rev} we review some basic aspects of the CFT approach to Kondo effect.
We especially focus on the role of the $g$-factor, and the low temperature scaling behavior of specific heat, susceptibility.
In Sec.~\ref{sec:QCD-Kondo} we perform the CFT analysis for QCD Kondo effect.
In Sec.~\ref{sec:in-medium} we first study the (1+1)-dimensional effective field theory for the finite density system with clarifying its specific symmetry.
We then compute the $g$-factor for this case, and find that the factor is irrational. This implies that the corresponding IR fixed point of QCD Kondo effect is described as the non-Fermi liquid.
In Sec.~\ref{sec:mag-system} we apply the CFT approach to the strong magnetic field system using a similar (1+1)-dimensional effective theory.
We show that the large $N$ analysis is naturally applicable to the situation in the strong magnetic field limit, due to the color symmetry enhancement.
In Sec.~\ref{sec:discussion} we conclude this paper with some remarks and discussions.

\section{CFT approach to $k$-channel SU($N$) Kondo effect}
\label{sec:CFT_rev}

We briefly summarize the CFT approach to multi-channel SU($N$) Kondo effect
(See~\cite{Affleck:1990iv} and \cite{Kimura:2016zyv} for more details).
The Kondo model is originally a three-dimensional model of the bulk fermions interacting with the localized impurity.
Under the assumption such that the impurity is sufficiently dilute, we consider the $s$-wave approximation, which ends up with the effective one-dimensional system along the radial direction.
The effective one-dimensional Hamiltonian of the $k$-channel SU($N$) Kondo model is given by
\beq
H
&=& H_{0} + H_{\rm{K}},
\label{TotalHmiltonian}
\eeq
where the free fermion Hamiltonian is
\beq
H_{0}
&=& \int^{\infty}_{0} dx \left[ i \psi_{L}^{\dagger} (x) \frac{\partial \psi_{L}(x) }{ \partial x } - i \psi_{R}^{\dagger}(x) \frac{ \partial \psi_{R}(x) }{ \partial x } \right] ,
\label{FreePart}
\eeq
and the Kondo interaction term is
\beq
H_{\rm{K}}
&=& \frac{ \lambda_{\rm{K}} }{4} S^{a} \left( \psi_{L}^{\dagger} (0) + \psi_{R}^{\dagger} (0) \right) t^{a} \left( \psi_{L}(0) + \psi_{R} (0) \right). 
\label{KondoInteraction}
\eeq
Here the one-dimensional coordinate $x$ is the relative coordinate between the bulk fermion and the impurity.
$\psi_{L}$ and $\psi_{R}$ are left and right moving fermions.
These fermions have SU($N$) spin and SU($k$) flavor degrees of freedom, but here we have suppressed those indices in Eqs. (\ref{FreePart}) and (\ref{KondoInteraction}).
$S^{a} \, (a=1, 2, \cdots, N^{2}-1)$ is a localized SU($N$) spin at the origin, and $t^{a}$ is the generators of SU($N$) group.
Since the bulk fermions satisfy the boundary condition $\psi_{L}(x) = \psi_{R}(-x)$, the Hamiltonian (\ref{TotalHmiltonian}) can be rewritten as
\beq
H
&=& \int^{\infty}_{-\infty} dx \left[ i \psi_{L}^{\dagger}(x) \frac{\partial \psi_{L}(x) }{ \partial x}  + \lambda_{\rm{K}} S^{a} \psi_{L}^{\dagger} (x) t^{a} \psi_{L} (x) \delta(x) \right].
\label{ModifiedHamiltonian}
\eeq
Below, we will suppressed the subscript $L$.
In terms of the spin, flavor and charge currents,
\begin{subequations}\label{eq:currents}
\beq
J^{a} (x)
&=& : \psi^{\dagger}(x) t^{a} \psi (x):, \label{ColorCurrent} \\
J^{A} (x)
&=& : \psi^{\dagger} (x) T^{A} \psi(x):, \label{FlavorCurrent} \\
J (x)
&=& : \psi^{\dagger} (x) \psi (x) :, \label{ChargeCurrent}
\eeq
\end{subequations}
we can express the Hamiltonian (\ref{ModifiedHamiltonian}) in the Sugawara form
\beq
H
&=& \int dx \, \left[ \frac{ 1 }{ N + k } J^{a} (x) J^{a} (x) + \frac{ 1 }{ k + N } J^{A}(x) J^{A} (x) + \frac{ 1 }{ 2 N k } J(x) J(x) + \lambda_{\rm{K}} J^{a} S^{a} \delta(x) \right] . \nonumber \\
\eeq
In Eqs.~\eqref{eq:currents}, $: O(x) O(x) : = \lim_{\epsilon \to 0} \left\{ O(x) O(x + \epsilon ) - \langle O (x) O(x + \epsilon ) \rangle \right\}$ is the normal order product, which subtracts the singular part at $\epsilon \to 0$. $T^{A} \, ( A = 1, 2, \cdots, k^{2} - 1)$ in Eq.~(\ref{FlavorCurrent}) is the generator of the flavor SU($k$) group.
Redefining the color current as
\beq
\mathcal{J}^{a} (x)
&=& J^{a} (x) + \frac{ \lambda_{\rm{K}} }{ 2 (N + k)} S^{a} \delta(x),
\eeq
we can rewrite the Hamiltonian as
\beq
H
&=& \int dx \, \left[ \frac{ 1 }{ N + k } \mathcal{J}^{a} (x) \mathcal{J}^{a} (x) + \frac{ 1 }{ k + N } J^{A}(x) J^{A} (x) + \frac{ 1 }{ 2 N k } J(x) J(x) \right] ,
\eeq
up to a constant term which does not contain the fermion fields.
We see the Hamiltonian is separated into the color, flavor and charge parts, respectively.
Furthermore, the effect of the impurity is reflected on the boundary of the theory.

Equivalently, the above effective (1+1)-dimensional theory can be expressed by the Wess--Zumino--Witten (WZW) model (See, for example, \cite{DiFrancesco:1997nk,Abdalla:2001NPM})
\begin{align}
 S
 = S_{k} [ g \in {\rm{SU}}(N) ] + S_{N} [ h \in {\rm{SU}}(k) ] + \frac{Nk}{2} \int d^{2}x \, (\partial_{\mu} \phi )^{2},
\label{WZWmodel}
\end{align}
where the WZW action $S_{n} [g]$ is given by
\begin{align}
S_{n} [g]
 = \frac{n}{16 \pi } \int d^{2} x\, \partial_{\mu} g \partial^{\mu} g^{-1} + \frac{n}{24 \pi } \int d^{3} x \, \epsilon^{\mu \nu \lambda }
 (g^{-1} \partial_{\mu} g) (g^{-1} \partial_{\nu} g) (g^{-1} \partial_{\lambda} g).
 \label{WZWaction}
\end{align}
In Eq.~(\ref{WZWmodel}), $g$ ($h$) is an element of SU($N$) (SU($k$)) group, while $k$ ($N$) is the level of the WZW action.
The bosonic field $\phi(x)$ describes the U(1) degrees of freedom associated with the charge current \eqref{ChargeCurrent}.
In particular, $k$ corresponds to the number of channel (flavor).
The WZW model (\ref{WZWmodel}) is also separated into the spin, flavor, and charge parts with the symmetry
\begin{align}
 \widehat{\SU(N)}_k \times \widehat{\SU(k)}_N \times
 \widehat{\U(1)}_{Nk}
 \, .
 \label{eq:sym_Kondo}
\end{align}
This factorization reflects the spin-charge separation in the effective one-dimensional theory, and $\widehat{G}_k$ is associated with the Kac--Moody algebra at level $k$.

In the CFT approach, it is essential to specify the symmetries of the Kondo system as well as the the representation of the impurity $R_\text{imp}$, e.g. $s$-spin representation for $\SU(2)$-spin interaction.
The fundamental parameters $(N,k,R_\text{imp})$ characterize the properties of the Kondo system near the IR fixed point, and the analysis based on CFT can be performed with them.
The method using CFT allows us to evaluate several quantities explicitly.
For example, one can compute the free energy, which characterizes thermodynamic properties of the system,
\begin{align}
 F = L f_\text{bulk} + f_\text{imp}
 \label{eq:free_en}
\end{align}
where the first term is the bulk contribution with the system size $L$ and the second one is the contribution of the impurity.
We are in particular interested in the impurity contribution to see the specific behavior under the Kondo effect.
What we focus on in this paper are:
\begin{enumerate}
 \item Boundary entropy ($g$-factor) at zero temperature 
 \item Specific heat \& susceptibility at low temperature 
\end{enumerate}
These quantities are exactly computed using the boundary CFT and the CFT perturbation theory with respect to the leading irrelevant operator, which are characterized by the fundamental parameters $(N,k,R_\text{imp})$ of the Kondo system.
In this paper we refer to our previous work~\cite{Kimura:2016zyv} for the formulae in the $\SU(N)$ Kondo model.

\subsection{Boundary entropy: $g$-factor}

In the Kondo system, the thermodynamic entropy can contain the contribution from the impurity associated with the free energy $f_\text{imp}$ \eqref{eq:free_en}, that can be exactly computed~\cite{Affleck:1991tk}
\begin{align}
 S_\text{imp}
 & =
 \log g(R_\text{imp})
\end{align}
where $g(R_\text{imp})$ depending on the impurity representation $R_\text{imp}$ is called the $g$-factor, which counts the degeneracy of the residual impurity spin at the IR fixed point, and monotonically decreases in the renormalization flow from UV to IR.
Furthermore, this is a useful quantity to see whether the IR fixed point is described as the Fermi liquid or the non-Fermi liquid:
If the $g$-factor shows an integer value, the system is described as the Fermi liquid, while it is the non-Fermi liquid for non-integer $g$, typically observed in the overscreening Kondo system.

For latter convenience, let us show the formula for the fundamental $\textbf{N}$ and antifundamental $\overline{\textbf{N}}$ representations
\begin{align}
 g(\textbf{N}) = g(\overline{\textbf{N}})
 & =
 q^{(N-1)/2} + q^{(N-3)/2} + \cdots + q^{-(N-1)/2}
 \label{eq:q-dim_fund}
\end{align}
where $q=\exp\left(2\pi i/(N+k)\right)$ for $k$-channel $\SU(N)$ Kondo system.
See, for example, \cite{Kimura:2016zyv} for derivation.

\subsection{Low temperature behavior}
\label{sec:low-T}

In addition to the quantity at the IR fixed point, we can exactly compute the low temperature scaling behavior of the specific heat and susceptibility, based on the conformal perturbation theory.
First of all, the bulk contribution depends only on the total central charge for the specific heat~\cite{Bloete:1986qm,Affleck:1986bv}
\begin{align}
 C_\text{bulk}
 & =
 \frac{\pi}{3} \, c \, T
 \label{eq:C_bulk} 
\end{align}
where $c = Nk$ is for the $k$-channel $\SU(N)$ Kondo system, while it depends on the channel number for the susceptibility~\cite{Affleck:1986sc}
\begin{align}
 \chi_\text{bulk} & = \frac{k}{2\pi}
 \, .
 \label{eq:chi_bulk} 
\end{align}

The impurity contribution exhibits an interesting dependence on the fundamental parameters $(N,k)$.
The computation with the WZW model for the single-channel system $(k=1)$ yields~\cite{Affleck:1990zd}
\begin{align}
 C_\text{imp}
 =
 - \lambda_1 \frac{k(N^2-1)}{3} \pi^2 T
 \qquad
 \chi_\text{imp}
 =
 - \lambda_1 \frac{k (N+k)}{2}
 \, .
\end{align}
Then the multi-channel system $(k>1)$ shows~\cite{Kimura:2016zyv}
\begin{align}
 C_\text{imp}
 &
 =
 \begin{cases}\displaystyle
  \frac{ \lambda^{2} }{2} \pi^{1 + 2 \Delta} ( 2\Delta)^{2} (N^{2}-1) (N + k / 2 ) \left[ \frac{ 1 - 2 \Delta }{ 2 } \right] \frac{ \Gamma(1/2 - \Delta ) \Gamma(1/2) }{ \Gamma(1-\Delta ) } \, T^{2 \Delta} & (k > N) \\[1em]
  \displaystyle
  \lambda^{2} \pi^{1 + 2 \Delta} ( N^{2}-1 )( N + k/2 ) ( 2\Delta )^{2} \, T \, {\rm{log}} \left( \frac{T_{\rm{K}}}{T} \right) & (k = N) \\[1em]
    \displaystyle
  - \lambda_1 \frac{k(N^2 - 1)}{3} \pi^2 T
  + 2 \lambda^{2} \pi^{2} (N^{2}-1) (N + k/2) \frac{ 2 \Delta }{ 1 + 2 \Delta } \left( \frac{\beta_{\rm{K}}^{-2 \Delta+1} }{ 2 \Delta -1 } \right)T & (N > k > 1)
 \end{cases}
 \, ,
 \label{eq:C_imp}
\end{align}
\begin{align}
 \chi_\text{imp}
 & =
 \begin{cases}\displaystyle
  \frac{\lambda^2}{2} \pi^{2\Delta -1 } (N + k/2 )^{2} (1-2\Delta) \frac{ \Gamma(1/2 - \Delta ) \Gamma(1/2) }{ \Gamma(1-\Delta ) } \, T^{2\Delta - 1} & (k > N) \\[1em] \displaystyle
  2 \lambda^2 (N + k/2 )^{2} \, \log \left(\frac{T_\text{K}}{T}\right) & (k = N) \\[1em] \displaystyle
  - \lambda_1 \frac{k(N+k)}{2}
  + 2 \lambda^{2} (N + k/2)^{2} \left( \frac{ \beta_{\rm{K}}^{-2 \Delta + 1} }{ 2 \Delta -1 } \right) & (N > k > 1)
 \end{cases}
 \, ,
 \label{eq:chi_imp}
\end{align}
where $\beta_{\rm{K}} = 1 / T_{\rm{K}}$ and $\Delta = N/(N+k)$.
Two unknown parameters $\lambda_1$ and $\lambda$ are coupling constants in the leading irrelevant operators with which the CFT perturbation is applied:
\begin{subequations}\label{eq:pert_op}
 \begin{align}
  \mathcal{O} & = \lambda \mathcal{J}_{-1}^a \phi^a (x)
  \, , \\
  \mathcal{O}_1 & = \lambda_1 \mathcal{J}^a \mathcal{J}^a(x)
  \, .
 \end{align}
\end{subequations}
See Refs.~\cite{Affleck:1990iv, Kimura:2016zyv} for details.
In the regime $N \le k$, we can focus on the operator with the coupling $\lambda$, which is specific to the non-Fermi liquid case, since the leading order contribution of the coupling $\lambda$ is dominant in this case.
In the regime $N > k > 1$, however, we have to consider another operator with the coupling $\lambda_1$ which is for the Fermi liquid since these contributions are in the same order.
This peculiar behavior in this regime $N > k > 1$ is called the Fermi/non-Fermi mixing~\cite{Kimura:2016zyv}, and may affect the universality of the Wilson ratio as explained below.
The scaling behaviors are summarized as
\begin{align}
 C_\text{imp}
 \propto
 \begin{cases}
  T^{2N/(N+k)} & (k > N) \\
  T \log \left( T_\text{K}/T \right) & (k = N) \\
  T & (N > k > 1)
 \end{cases}
 \qquad
 \chi_\text{imp}
 \propto
 \begin{cases}
  T^{(N-k)/(N+k)} & (k > N) \\
  \log(T_\text{K}/T) & (k = N) \\
  \text{const.} & (N > k > 1)
 \end{cases}
\end{align}

In general, the impurity contributions to the specific heat and susceptibility contain the coupling constants $\lambda$ and $\lambda_{1}$ which depend on the microscopic details of the system, and thus not universal.
Wilson showed that such a non-universal dependence can be canceled in a specific ratio of the specific heat and susceptibility, which is called the Wilson ratio~\cite{Wilson:1974mb}
\begin{align}
 R_\text{W}
 & =
 \left( \frac{\chi_\text{imp}}{C_\text{imp}} \right)
 \Bigg/
 \left( \frac{\chi_\text{bulk}}{C_\text{bulk}} \right)
 \, .
\end{align}
For the single-channel system $(k=1)$, it is given by
\begin{align}
 R_\text{W}
 =
 \frac{N}{N-1}
 \, .
\end{align}
For the multi-channel system $(k>1)$, there are two possibilities.
In the regime $N < k$, it becomes a universal constant which depends only on the fundamental parameters $(N,k)$~\cite{Ludwig:1994nf}
\begin{align}
 R_\text{W}
 & =
 \frac{(N+k/2)(N+k)^2}{3N(N^2-1)}
 \qquad
 (N < k)
 \, ,
\end{align}
while, in the regime $N > k > 1$, it fails to cancel the non-universal factor~\cite{Kimura:2016zyv}
\begin{align}
 R_\text{W}
 & =
 \frac{(N+k/2)(N+k/3)}{N^2-1}
 \frac{\displaystyle \gamma - \frac{k(N+k)}{(N+k/2)^2}}
      {\displaystyle \gamma - \frac{k(N+k/3)}{N (N+k/2)}} 
 \qquad
 (N > k > 1)
\end{align}
where the dimensionless constant is defined
\begin{align}
 \gamma
 & =
 4 \, T_\text{K}^{2\Delta-1} \, \frac{\lambda^2}{\lambda_1} 
 \, .
\end{align}
Therefore, in this case, the Wilson ratio does depend on the microscopic details of the system, and then it is no longer universal.

\section{QCD Kondo effect: CFT analysis}
\label{sec:QCD-Kondo}

\begin{figure}
\begin{center}
\includegraphics[width=0.7 \textwidth]{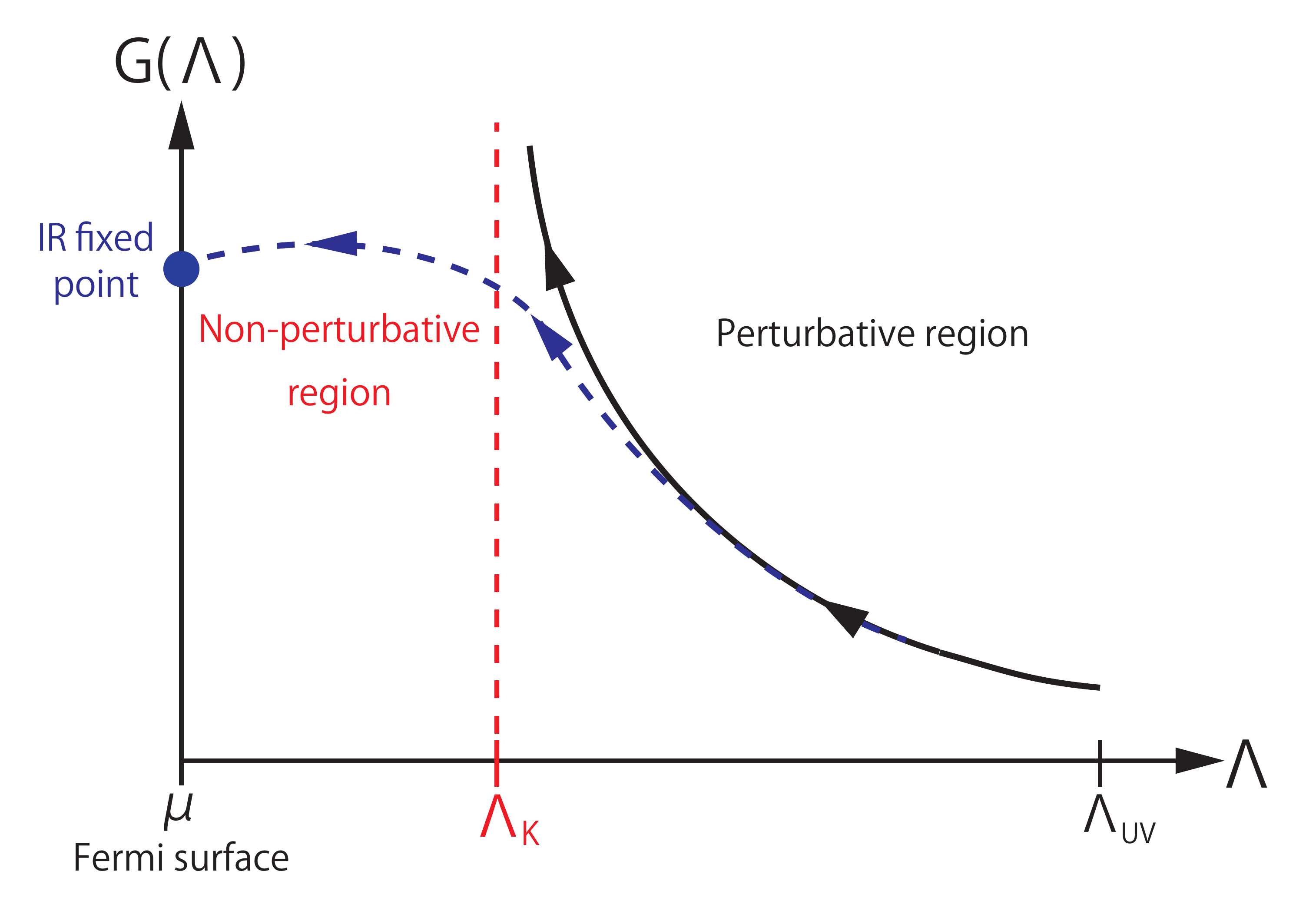}
\end{center}
\caption{
Schematic picture of the flow of the effective interaction $G(\Lambda)$.
The black solid line is a perturbative flow of $G(\Lambda)$, while the blue dashed line is a non-perturbative flow. $\Lambda_\text{K}$ stands for the Kondo scale.
 }
 \label{fig:scale_dependence}
\end{figure}

The QCD Kondo effect is the Kondo effect induced by the color exchange interaction between light quarks near the Fermi surface and a heavy quark impurity.
One of the present authors and others have studied the QCD Kondo effect at finite chemical potential and in strong magnetic fields in terms of the perturbative renormalization group approach \cite{Hattori:2015hka,Ozaki:2015sya}.
Then, we find the Kondo scale $\Lambda_\text{K}$, at which the effective interaction $G(\Lambda)$ between the light quark and the heavy quark impurity diverges.
The resultant expression of the Kondo scale is given by~\cite{Hattori:2015hka,Ozaki:2015sya}
\beq
\Lambda_{\rm{K}}
\sim \Lambda_{0} \, {\rm{exp}} \left( - \frac{ 4\pi }{ N_{c} \alpha_{s} {\rm{log}} \left(  \frac{4\pi}{ \alpha_{s} } \right) } \right),
\eeq
where $\Lambda_{0} = \mu$ for finite density system, and $\Lambda_{0} = \sqrt{eB}$ for strong magnetic field system.
In Fig.~\ref{fig:scale_dependence}, the schematic picture of the scale dependence of $G(\Lambda)$ is shown.
At the finite temperature, the IR cut-off is replaced by the temperature $T$. Then, the typical scale of the Kondo system is given by the Kondo temperature $T_{\rm{K}}$, instead of $\Lambda_{\rm{K}}$.
These two scales are the same order~\cite{Yasui:2017bey}.
Below the Kondo temperature or Kondo scale, the system would be in the non-perturbative regime, and thus the perturbative approach is no longer valid.
In this section, we apply the CFT approach to investigate the QCD Kondo effect around the IR fixed point.
We will obtain several observables including the specific heat, and the color susceptibility.

Furthermore, we study the QCD Kondo effect with multi-flavors.
Although the multi-channel Kondo problem has a rich structure which leads to the non-Fermi liquid behavior, there are some difficulties of realizing such a multi-channel system in condensed matter physics because of the fine-tuning of channel degeneracy.
QCD Kondo effect could overcome this difficulty of realizing the multi-channel Kondo effect because we can naturally introduce flavor degrees of freedom to quarks.
For the QCD Kondo system, we consider the sufficiently large chemical potential $\mu$ or strong magnetic field $\sqrt{eB}$. 
Comparing with these energy scales, the current quark masses of $u$, $d$, and $s$ are sufficiently small, and thus we can take into account the corresponding flavor symmetry for light quarks.

\subsection{Finite density system}
\label{sec:in-medium}

In a finite density system, we have the Fermi surface of light quarks.
Assuming the Fermi surface is spherically symmetric in the momentum space, we can apply the $s$-wave approximation, which allows us to consider the (1+1)-dimensional setup.
Then the massless excitation near the Fermi surface is actually given by a momentum fluctuation perpendicular to the surface, which is effectively described by (1+1)-dimensional WZW model.

\subsubsection{Effective field theory and symmetry in (1+1)-dimensions}

In this study, we take the chemical potential $\mu$ much larger than the QCD scale: $\mu \gg \Lambda_{\rm{QCD}}$ so that the gauge coupling $g_{s}$ as well as $\alpha_{s} = g_{s}^{2} / 4\pi$ are sufficiently small. Integrating the gluon fields, the QCD action at high densities becomes
\beq
S^\text{4D}
&=& \int d^{4}x \,  \bar{\psi} (i\gamma^{\mu} \partial_{\mu}  + \mu \gamma^{0} ) \psi \nonumber \\
&& + \, 4\pi \alpha_{s} \int d^{4}x d^{4} y \, \bar{\psi} \gamma^{\mu} t^{a} \psi (x)
D_{\mu \nu}^{ab} (x, y; \mu) \bar{Q} \gamma^{\nu} t^{b} Q (y)
+ O(\alpha_{s}^{2}),
\label{eq:4d_action}
\eeq
where $(\psi, \bar{\psi})$ is the light quark field with $N_{f}$ flavors while $(Q,\bar{Q})$ is a single heavy quark field.
$D_{\mu \nu}^{ab} (x, y; \mu)$ stands for the gluon propagator at finite density.
Since the gauge coupling is sufficiently small, we neglect higher order terms with respect to $\alpha_{s}$.
Furthermore, considering the heavy quark mass limit, the vertex of the heavy quark impurity leads to
$ \bar{Q} \gamma^{\nu} Q \sim \bar{Q} \gamma^{0} Q = Q^{\dagger} Q$ and $Q^{\dagger} t^{a} Q(x)
\to C^{a} \delta(x)$, which is the spatially localized color charge with the (anti-)fundamental representation of the heavy quark impurity.
Then, the action reads
\beq
S^\text{4D}
&=& \int d^{4}x \, \bar{\psi} (i \gamma^{\mu} \partial_{\mu} + \mu \gamma^{0} )\psi
+ 4\pi \alpha_{s} \int d^4 x d^{4} y \, \bar{\psi} \gamma^{0} t^{a} \psi (x)
D_{0 0}^{ab} (x, y; \mu) C^{b} \delta(y)
\eeq
In the finite density, the gluon propagation is suppressed by the Tomas--Fermi screening with the screening mass $m_{g}^{2} = (\alpha_{s} / \pi) N_{f} \mu^{2}$.
Consequently, the interaction between the light quark and the heavy quark impurity becomes the $\delta$-function type interaction. This enables us to apply the $s$-wave approximation. 
Under the $s$-wave approximation near the Fermi surface, the (1+1)-dimensional effective action of high density QCD in the presence of the heavy quark impurity can be written as~\cite{Shuster:1999tn, Kojo:2009ha}
\beq
S_\text{eff}^\text{2D}
&=& \int d^{2} x \left[ \bar{\Psi} \left[ i \Gamma^{\mu} \partial_{\mu}  \right] \Psi 
+ G \, \Psi^{\dagger} t^{a} \Psi C^{a} \delta(x) \right],
\label{SUN_Kondo}
\eeq
where the two dimensional Dirac matrices are given by $\Gamma^{0} = \sigma^{1}$, $\Gamma^{z} = -i \sigma^{2}$.
We choose the $z$-direction to be perpendicular to the Fermi surface.
$\Psi$ is a two component quark field with $2N_{F}$ flavors.
Here the factor $2$ of the flavor comes from the spin degrees of freedom in (3+1)-dimensions.
Since there is no rotational symmetry in (1+1)-dimensional system, there is no spin.
Therefore, the spin degrees of freedom in (3+1)-dimension enlarge the flavor symmetry in (1+1)-dimensional system.
The relation between the two component quark field $\Psi$ in (1+1) dimensions and the ordinary four component quark field $\psi$ with $N_{F}$ flavors in (3+1)-dimensions is the following:
The two component quark field $\Psi$ with $2N_{f}$ flavors is defined by
\beq
\Psi 
&=& \left[
\begin{matrix}
{\rm{e}}^{i \mu z \Gamma^{5} } \varphi_{\uparrow}  \\
{\rm{e}}^{i \mu z \Gamma^{5} } \varphi_{\downarrow}  \\
\end{matrix}
\right],
\eeq
where $\Gamma^{5} = \Gamma^{0} \Gamma^{z} = \sigma^{3}$, and $\varphi_{\uparrow}^\text{T} = ( \psi_{R + }, \psi_{L -})$, $\varphi_{\downarrow}^\text{T} = ( \psi_{L+}, \psi_{R-})$. 
$\psi_{R, L}$ and $\psi_{+, -}$ are quark fields in the chiral and spin bases in (3+1)-dimensions. 
The dimensionless coupling $G$ corresponds to $\lambda_{K}$ in Sec.~\ref{sec:CFT_rev}.
Therefore, the effective action (\ref{SUN_Kondo}) can be regraded as a (1+1) dimensional $k$-channel $SU(N_{c})$ Kondo model with $k = 2N_{f}$.

In the perturbative regime, the interaction term in Eq. (\ref{SUN_Kondo}) is induced by the $s$-wave projected one-gluon exchange between light quarks and the heavy quark. Near the Fermi surface, the coupling $G$ is obtained by~\cite{Evans:1998ek}
\beq
G
&=& \rho_{F} \int d \Omega_{q} \frac{ (ig_{s})^{2} }{ q^{2} - m_{g}^{2} } \nonumber \\
&=& \frac{\mu^{2}}{(2\pi)^{2}} \int d\Omega_{q} \frac{ (ig_{s})^{2} }{ - 2 \mu^{2} (1 - {\rm{cos}} \theta) - m_{g}^{2} } \nonumber \\
&=& \alpha_{s} \, {\rm{log}} \left( \frac{ 4 \mu^{2} }{ m_{g}^{2} } \right), 
\label{coupling1}
\eeq
where $\rho_{F}= \mu^{2} / (2\pi)^{2}$ is the density of state on the Fermi surface.%
\footnote{Here we only consider the color electric interaction since the color magnetic interaction is suppressed by $1/M_{Q}$ with large heavy quark mass limit.}
Then, by using the perturbation theory of the effective action (\ref{SUN_Kondo}) with respect to the coupling (\ref{coupling1}), one can reproduce the perturbative result of QCD Kondo effect obtained in~\cite{Hattori:2015hka}.%
\footnote{In Ref.~\cite{Hattori:2015hka}, the gluon propagator at finite density is just taken as $-1/m_{g}^{2}$ with $q \to 0$. However, depending on the angle of the scattered light quark, the momentum transfer $q$ can be of the order of $\mu$ ($\gg m_{g}$). Taking into account the angle dependence on the gluon propagator, the authors of~\cite{Evans:1998ek} showed that the effective coupling has a logarithmic form as in (\ref{coupling1}). Once we use the same gluon propagator, we can reproduce the result of Ref.~\cite{Hattori:2015hka} from the effective action (\ref{SUN_Kondo}).}

Now, as mentioned in the beginning of this section, the perturbative approach is no longer valid below the Kondo scale.
To investigate QCD Kondo effect near the IR fixed point where the system is highly non-perturbative, we apply the CFT approach to the effective action (\ref{SUN_Kondo}) in the same way as discussed in Section~\ref{sec:CFT_rev}. 
We can express the effective Hamiltonian at the IR fixed point in the Sugawara form as
\begin{align}
 H
 & = \int dx \, \left[ \frac{ 1 }{ N_{c} + 2N_{f} }\mathcal{J}^{a} (x)\mathcal{J}^{a}(x) + \frac{1}{ 2N_{f} + N_{c} } J^{A} (x) J^{A}(x) + \frac{1}{ 4 N_{c} N_{f} } J(x) J(x) \right], 
\label{FiniteDensityHamiltonian}
\end{align}
where
\begin{subequations}
\beq
\mathcal{J}^{a}(x)
&=& J^{a}(x) + \frac{ G }{ 2 ( N_{c} + 2 N_{f} ) } C^{a} \delta(x)
\eeq
with the color, flavor and charge currents
\beq
J^{a}(x)
&=& : \Psi^{\dagger} (x) t^{a} \Psi(x) :, \\
J^{A}(x)
&=& : \Psi^{\dagger} (x) T^{A} \Psi(x) :, \\
J(x)
&=&  : \Psi^{\dagger} (x) \Psi(x) :,
\eeq
\end{subequations}
respectively. 
Here $t^{a} \ (a = 1, 2, \ldots, N_{c}^{2} - 1)$ is the generator of the color SU($N_{c}$) group, while $T^{A} \ (A = 1, 2, \ldots, (2N_{f})^{2} - 1)$ is the generator of the flavor SU($2N_{f}$) group.
The effective Hamiltonian (\ref{FiniteDensityHamiltonian}) is equivalent to the following WZW model with the definition \eqref{WZWaction}:
\beq
S_{2N_{f}} ( g \in {\rm{SU}}(N_{c} ) ) + S_{N_{c}} ( h \in {\rm{SU}} (2N_{f}) ) +  N_{c} N_{f} \int d^{2}x \, ( \partial_{\mu} \phi )^{2} ,
\label{FiniteDensityWZW}
\eeq
See, for example, Ref.~\cite{Abdalla:2001NPM} for the non-Abelian bosonization scheme.
We remark that a similar WZW action for the finite density QCD is obtained in Ref.~\cite{Kojo:2009ha}.
From the effective Hamiltonian (\ref{FiniteDensityHamiltonian}) or the chiral WZW model (\ref{FiniteDensityWZW}), we can read off the symmetry of the (1+1)-dimensional effective theory as 
\begin{align}
 \widehat{\SU(N_c)}_{2N_f} \times \widehat{\SU(2N_f)}_{N_c} \times
 \widehat{\U(1)}_{2N_c N_f}
 \, ,
 \label{eq:sym_fin}
\end{align}
which is equivalent to the symmetry of the $k$-channel $\SU(N)$ Kondo model~\eqref{eq:sym_Kondo} under the replacement $(N,k) \to (N_c, 2N_f)$.
The first and third factors are due to color $\SU(N_c)$ and baryon number $\U(1)$ symmetries.
The remaining one is of the $\SU(N_f)$ flavor symmetry, which is now enhanced to $\widehat{\SU(2N_f)}$.
This is because, during the dimensional reduction from the original (3+1)-dimensional theory to the (1+1)-dimensional theory, the number of spinor components is reduced as $4 \to 2$, and we have up and down spin states in the effective model.
We need to combine the flavor symmetry with this spin rotation symmetry.
The reason for the level $2N_f$ of $\widehat{\SU(N_c)}_{2N_f}$ is actually the same.
We remark that a similar argument on the flavor symmetry enhanced by the chirality has been provided in~\cite{Kanazawa:2016ihl}.

Although we have three kinds of degrees of freedom for the effective (1+1)-dimensional theory, the relevant part to our purpose is only color degrees of freedom.
First of all, to exhibit the Kondo effect, we need to take into account the non-Abelian interaction with the impurity.
In addition, as mentioned above, we have the up and down spin states in the effective model.
Nevertheless, we do not need to incorporate this spin degrees of freedom as a non-Abelian property of the interaction because the impurity spin is fixed in the heavy quark limit.
Thus the spin interaction with the impurity can be now negligible, and we can focus only on the color degrees of freedom $\widehat{\SU(N_c)}_{2N_f}$ in this case, which is the same as the ordinary Kondo problem.
As shown in eq.~\eqref{eq:sym_Kondo}, the Kondo problem is characterized by the fundamental parameters of spin, channel, and the impurity representation $(N,k,R_\text{imp})$.
In the case of QCD Kondo effect, we first apply $(N,k) = (N_c, 2N_f)$, and it is natural to assign the (anti-)fundamental representation to the heavy quark impurity, $R_\text{imp} = \mathbf{N}$ or $\bar{\mathbf{N}}$.

\subsubsection{IR behaviors of QCD Kondo effect}

\begin{table}[t]
 \begin{center}
   \begin{tabular}{@{\hspace{2em}}c@{\hspace{6em}}l@{\hspace{1.5em}}} \hline\hline
   $N_f$ & $g$-factor \\ \hline
    1 & $1.61803...$ \\
    2 & $2.24698...$ \\
    3 & $2.53209...$ \\ \hline\hline
   \end{tabular}
 \end{center}
 \caption{The $g$-factor of $\SU(3)$-color theory for $N_f = 1, 2, 3$ with $R_\text{imp} = \mathbf{3}$ ($\bar{\mathbf{3}}$) given by the formula~\eqref{eq:q-dim_SU3}. The asymptotic behavior of the $g$ factor is $g \to 3$ as $N_f \to \infty$.}
 \label{tab:g-fac}
\end{table}

Let us first compute the $g$-factor with the formula \eqref{eq:q-dim_fund}, and put $N_c = 3$.
The $\SU(3)$ impurity in the (anti-)fundamental representation $R_\text{imp} = \mathbf{3}$ ($\bar{\mathbf{3}}$) gives rise to
\begin{align}
 g
 & =
 q + 1 + q^{-1}
 = 1 + 2 \cos \left( \frac{2\pi}{3+ 2N_f} \right)
 \, .
 \label{eq:q-dim_SU3}
\end{align}
where $q = \exp \left( 2 \pi i /(3+2N_f)\right)$.
Now the $g$-factor is the same for the fundamental and anti-fundamental representations.
Table~\ref{tab:g-fac} shows this $g$-factor with $N_f = 1, 2, 3$.
We observe that the asymptotic behavior of the $g$ factor is $g \to 3$ as $N_f \to \infty$.
This is just the dimension of $\mathbf{3}$ ($\bar{\mathbf{3}}$) representation of $\SU(3)$ in a usual sense.
A remarkable point to this QCD Kondo effect is that we obtain an irrational $g$-factor even for $N_f = 1$ theory.%
\footnote{%
In particular, the $g$-factor with $N_f = 1$ is given by the golden ratio
\begin{align}
 1 + 2 \cos \left( \frac{2 \pi}{5} \right) = \frac{1 + \sqrt{5}}{2}
 \, ,
\end{align}
which also appears in the three-channel $\SU(2)$ Kondo system, described by $\widehat{\SU(2)}_3$ theory.
This connection is due to the level-rank duality of $\widehat{\SU(3)_2}$ and $\widehat{\SU(2)}_3$.
}
In other words, the QCD Kondo effect is always over-screening; cannot be critical nor under-screening.
This is due to the spin degrees of freedom of quarks.
Even starting with a single quark, it splits into up and down spin states in (1+1)-dimensional effective theory, while the heavy quark spin is fixed in the impurity limit (the heavy quark limit).
This means that, in the finite density QCD Kondo effect, the minimal number of channels is two when $N_f = 1$, which leads to the over-screening state.
A naive Kondo singlet state is the $q\bar{Q}$-bound state, but the situation is not so simple in this case because this kind of bound state picture is based on the Fermi liquid description.
Our analysis suggests that the finite density QCD Kondo effect is always described as the non-Fermi liquid at the IR fixed point, and thus a screening process of color degrees of freedom becomes also non-trivial.


In addition to the $g$-factor, which characterizes the IR fixed point, we can compute low temperature dependences of several quantities just by applying the method shown in Sec.~\ref{sec:low-T}.
Since the finite density QCD Kondo effect shows always $k>1$, as discussed above, we straightforwardly apply the formulas for the multi-channel system to this case just by putting $(N,k) = (N_c, 2N_f)$.
The bulk contribution to the specific heat and the susceptibility are obtained as
\begin{align}
 C_\text{bulk}
 =
 \frac{2\pi}{3} N_c N_f T
 \qquad
 \chi_\text{bulk}
 =
 \frac{N_f}{\pi}
 \, .
\end{align}
respectively.
Then the low temperature scaling of the impurity part is given by
\begin{align}
 C_\text{imp}
 & \propto
 \begin{cases}
  T^{2N_c/(N_c+2N_f)} & (2N_f > N_c) \\
  T \log \left( T_\text{K}/T \right) & (2N_f = N_c) \\
  T & (N_c > 2N_f > 1)
 \end{cases}
 \\
 \chi_\text{imp}
 & \propto
 \begin{cases}
  T^{(N_c-2N_f)/(N_c+2N_f)} & (2N_f > N_c) \\
  \log(T_\text{K}/T) & (2N_f = N_c) \\
  \text{const.} & (N_c > 2N_f > 1)
 \end{cases}
\end{align}
For $N_c=3$, the low temperature behaviors of $N_f = 1$ and $N_f > 1$ are essentially different since the former case shows the Fermi liquid type dependence, while the latter exhibits the non-Fermi liquid type low temperature behavior.
Furthermore, the case with $N_f=1$ shows the Fermi/non-Fermi mixing since its $g$-factor implies that its IR fixed point is described as the non-Fermi liquid.

The Wilson ratio of the QCD Kondo effect can be obtained by taking the ratio of the specific heat and the susceptibility. In the cases of $N_{c} \le 2N_{f}$, we find
\beq
R_{\rm{W}}
&=& \frac{ (N_{c} + N_{f} ) ( N_{c} + 2 N_{f} )^{2} }{ 3 N_{c} (N^{2}_{c} - 1 ) }. \ \ \ \ \  (N_{c} \le 2N_{f})
\eeq
The Wilson ratio is universal for these cases.
On the other hand, in the cases of $N_{c} > 2 N_{f}$, the Wilson ratio becomes
\beq
R_{\rm{W}}
&=& \frac{ ( N_{c} + 2 N_{f} )( N_{c} + 2 N_{f} / 3 ) }{ N_{c}^{2} -1 }
\frac{ \gamma - \frac{ 2N_{f} ( N_{c} + 2 N_{f} ) }{ (N_{c} + N_{f})^{2} } }{ \gamma - \frac{ 2 N_{f} (N_{c} + 2 N_{f} / 3 ) }{ N_{c} + N_{f} ) } }, \ \ \ \ \ (N_{c} > 2N_{f})
\eeq
where the dimensionless constant is defined as
\beq
\gamma
&=& 4 T_{\rm{K}}^{2 \Delta - 1 } \frac{ \lambda^{2} }{ \lambda_{1} }.
\eeq
with $\Delta = N_{c} / ( N_{c} + 2N_{f} )$.
The couplings $(\lambda,\lambda_1)$ are of the perturbation operators \eqref{eq:pert_op}, which would be obtained from the higher order terms in the original (3+1)-dimensional QCD action \eqref{eq:4d_action}.

\subsection{Strong magnetic field system}
\label{sec:mag-system}

In this Section, we consider the magnetized quark matter in the presence of a heavy quark impurity with $\sqrt{eB} \gg \mu$.
In this case, the effect of the magnetic field dominates, and then the magnetically induced QCD Kondo effect occurs \cite{Ozaki:2015sya}.
However, we have to keep in mind that since $\mu \neq 0$, the Fermi surface is still there, and the IR fixed point exists on the Fermi surface.
This is one of the reasons why we need the CFT analysis to study this situation, while the sytem in the background magnetic field at zero density $\mu = 0$ can be studied, for example, based on the lattice QCD~\cite{Bali:2014kia}.

\subsubsection{Effective field theory and symmetry in (1+1)-dimensions}

Let us derive the (1+1)-dimensional effective field theory describing QCD Kondo effect in the strong magnetic field.
In strong magnetic fields, quarks with electric charges are in the Lowest Landau level (LLL) state.
With the different electric charges of the bulk quarks, $Q_{u} = 2/3$ and $Q_{d}=-1/3$, we can divide the flavor sector as $N_{f} \to N_{f}^{(u)} \oplus N_{f}^{(d)}$ where $N_{f}^{(u,d)}$ is the number of the flavor having the electric charge $Q_{u} (Q_{d})$.
Accordingly, in the LLL approximation, the bulk quark part of the 4-dimensional action reads
\beq
S_{\psi}^\text{4D}
&=& \int d^{4} x \ \bar{\psi}_\text{LLL}^{(u)} \left[ i \gamma^{\mu} ( \partial_{\mu} + ig_{s} A_{\mu}) + \mu \gamma^{0} \right] \psi_\text{LLL}^{(u)} \nonumber \\
&& + \ ( u \leftrightarrow d ).
\eeq
In the symmetric gauge, the LLL quark fields with the spin basis are given by 
\beq
\psi_\text{LLL}^{(u,d)} (x) 
&=& \left[
\begin{matrix}
 \displaystyle
 \sum_{l=0}^{N_{L}^{(u,d)}-1} \phi_{l}^{(u,d)} (x_{\perp}) c_{l, c}^{(u,d)} (x_{\parallel}) \\
 0 \\
 \end{matrix}
 \right],
 \eeq
where $(l,c)$ are indices of the angular momentum and the color, respectively.
$N_L^{(u,d)}$ is the degeneracy of the LLL given by $N_{L}^{(u,d)} = V_{\perp} |Q_{u, d}eB| / (2\pi)$ with the transverse volume $V_\perp$.
$\phi_{l}^{(u,d)}$ is the perpendicular pert of the wave function in the LLL, 
\beq
\phi_{l}^{(u,d)} (x_{\perp})
&=& \sqrt{ \frac{ |eQ_{u,d} B| }{ 2\pi l ! } } \left( \frac{ |Q_{u,d}eB| }{ 2 } \right)^{l/2} ( x + i y )^{l} {\rm{e}}^{ - \frac{ |eQ_{u,d}B| }{4} (x^{2} + y^{2} ) },
\eeq 
while the $c_{l,c}^{(u,d)}$ is a two component spinor with $N_{f}^{(u)}$ ($N_{f}^{(d)}$) flavors. 
Integrating the perpendicular parts of the coordinate, we can reduce the action to the (1+1)-dimensional one:
\beq
S_{\psi}^\text{2D}
&=& \int d^{2} x_{\parallel} \ \bar{c}_{l^{\prime}, c^{\prime} }^{(u)} \left[ i \Gamma^{\mu} ( \partial_{\mu} \delta_{l^{\prime}l} \delta_{c^{\prime}c} + ig_{s} [A_{\mu}^{(u)}]_{l^{\prime}, c^{\prime}; l, c} ) \right] c_{l,c}^{(u)} \nonumber \\
&& + \ (u \leftrightarrow d),
\label{2DactionStrongB}
\eeq
where the definition of the two dimensional Dirac matrices is the same as in the previous section,%
\footnote{In this study, we apply the magnetic field $B$ in $z$-direction.} and the chemical potential is absorbed into the field by shifting it as $c_{l,c}^{(u,d)} \to {\rm{e}}^{i \mu z \Gamma^{5} } c_{l,c}^{(u,d)}$.
The gauge field in Eq.~(\ref{2DactionStrongB}) is given by
\beq
[A_{\mu}^{(u,d)}]_{l^{\prime}, c^{\prime}; l, c} (x_{\parallel})
&=& \int d^{2} x_{\perp} \ \phi^{(u,d) *}_{l^{\prime}} (x_{\perp}) [A_{\mu}(x)]_{c^{\prime}, c} \phi_{l}^{(u,d)}(x_{\perp}).
\eeq
Now, the bosonized version of the the bulk quark part of the action can be expressed in terms of the WZW action as
\begin{align}
S_{\text{kin}}^\text{2D}
 & = S_{N_f^{(u)}} (\mathcal{G}^{(u)}) + S_{N_f^{(d)}} (\mathcal{G}^{(d)})
 \, ,
\label{WZW_B0}
\end{align}
with the definition \eqref{WZWaction}.
Here $\mathcal{G}^{(u)}$ and $\mathcal{G}^{(d)}$ are elements of $\SU(N_{L}^{(u)} N_{c})$ and $\SU(N_{L}^{(d)} N_{c})$ groups, respectively.
Precisely speaking, in addition to the action \eqref{WZW_B0}, there are flavor and charge parts similar to \eqref{WZWmodel}, which are irrelevant to the current case.
This is a generalization of bosonization scheme for the strong magnetic field system with a single flavor $N_{f}=1$~\cite{Hayata:2013sea} to the multi-flavor cases $N_{f} > 1$. 
In this study, we consider a simple case, namely, $N_{f}^{(u)} \neq 0$ and $N_{f}^{(d)} = 0$, and the heavy quark impurity has the charge $+2/3$.
Further generalized cases will be investigated in the future work.

We can read off the symmetry of the (1+1) dimensional effective theory of the strong magnetic field system from the WZW action 
\begin{align}
 \widehat{\SU(N_L^{(u)} N_c)}_{N_f^{(u)}} \times \widehat{\SU(N_f^{(u)})}_{N_L^{(u)} N_c} \times
 \widehat{\U(1)}_{N_L^{(u)} N_c N_f^{(u)}}
 \, .
 \label{eq:sym_mag}
\end{align}
If there are also $N_f^{(d)}$ $d$-type quarks with the charge $e_d = (-1/3)\times e$, there appears another system with the symmetry~\eqref{eq:sym_mag} by replacing $(N_L^{(u)}, N_c, N_f^{(u)}) \to (N_L^{(d)}, N_c, N_f^{(d)})$.
The $\SU(N_c)$-color symmetry is now enhanced to $\SU(N_L^{(u)} N_c)$ due to the LLL degeneracy, and thus the rank of this enhanced color symmetry $\SU(N_L^{(u)} N_c)$ becomes very large, which allows us to apply the large $N$ approximation to this case.
In contrast to the finite density system, there is no enhancement of the flavor symmetry.
This is because the LLL state is fully polarized, and thus the spin degree of freedom is frozen in the strong magnetic field.
Assuming the impurity heavy quark is in the (anti-)fundamental representation, the magnetically induced QCD Kondo effect is specified by the fundamental parameters $(N,k,R_\text{imp}) = (N_L^{(u)} N_c, N_f^{(u)}, \text{fund/anti-fund})$.
This assumption is justified when the magnetic field scale is sufficiently larger than the heavy quark mass, so that the LLL approximation is also applicable to the heavy quark.
Otherwise, the heavy quark does not behave under the $\SU(N_L^{(u)} N_c)$ transformation, but the ordinary $\SU(N_c)$ color transformation.
Similarly it is also possible to consider the situation where the light quark is $u$-type and the heavy quark is $d$-type, and vice versa.
In these cases, the light and heavy quarks belong to the different symmetry groups, and the analysis using the boundary CFT will be much more involved.
We would like to return to this problem in the future.

\subsubsection{IR behaviors of magnetically induced QCD Kondo effect}

In the following, we use $(N_L, N_c, N_f)$ suppressing $(u)$ for simplicity as long as no confusion.
To characterize the IR fixed point of the current Kondo problem, we compute the $g$-factor for the (anti-)fundamental representation 
for the strong magnetic field system, which is given by the formula~\eqref{eq:q-dim_fund} with $q = \exp \left( 2\pi i/(N_L N_c + N_f) \right)$.
Expanding this expression under the assumption $N_L \gg 1$ with $N_f$ fixed, we obtain
\begin{align}
 g & =
 N_f - \frac{N_f (N_f^2 - 1)}{(N_L N_c)^2} \frac{\pi^2}{6}
 + O \left( N_L^{-3} \right) 
 \, .
 \label{eq:g-fac_large-N}
\end{align}
In the large $N_L$ limit, corresponding to the large $B$ limit, the $g$-factor is approximated to $g = N_f$, and the correction is highly suppressed since it starts with $O \left( N_L^{-2} \right)$.
This (almost) integer behavior implies that the QCD Kondo effect is described as the Fermi liquid in the strong magnetic field limit, and thus the low-temperature scaling of the specific heat and so on is expected to exhibit the Fermi liquid behavior.

These results implying the Fermi liquid behavior at the IR fixed point also suggest that we correspondingly observe the low temperature behavior described by the Fermi liquid
\begin{align}
 C_\text{imp} \propto T,
 \qquad
 \chi_\text{imp} \propto \text{const}
 \, .
\end{align}
The Wilson ratio in this case is given by
\begin{align}
 R_\text{W}
 = \frac{N_L N_c}{N_L N_c - 1}
 \ \stackrel{N_L \to \infty}{\longrightarrow} \
 1
 \, .
\end{align}
Such a behavior is expected to be observed in the QCD Kondo effect occurring in the strong magnetic field, and would be a signal for it.

\section{Discussion}
\label{sec:discussion}

In this paper we have studied QCD Kondo effect both in finite density and strong magnetic field systems based on the CFT approach.
We have derived the (1+1)-dimensional WZW model, which is the effective theory of the QCD Kondo effect, and then pointed out that, in contrast to the ordinary Kondo effect, the spin interaction does not play any role in both QCD Kondo effects.
In both cases, we have observed the symmetry enhancement, which is indeed a specific feature of QCD Kondo effect since there is no such enhancement in the conventional Kondo effects.
In the finite density system, the flavor symmetry is enhanced due to the spinor structure, while the color symmetry is enhanced in the magnetic system due to the LLL degeneracy.

Due to the flavor symmetry enhancement for the finite density system, we have shown that the corresponding QCD Kondo effect is always over-screening, which exhibits the non-Fermi liquid behavior at the IR fixed point.
On the other hand, for the strong magnetic field system, we have performed the large $N$ analysis thanks to the enhancement of color symmetry.
We have obtained the Fermi liquid behavior in the strong magnetic field limit, and its correction just starts in $O(B^{-2})$.
Applying the CFT analysis to the QCD Kondo problem, we have shown the impurity contribution to the low-temperature dependence of specific heat, susceptibility, which could also distinguish the universality of QCD Kondo effects.

It is important to discuss several physical situations in which QCD Kondo effect can occur, and possible signatures of QCD Kondo effect in QCD phenomenology.
In the core of neutron star and magnetar, a high density quark matter would exist, as well as strong magnetic fields.
If the high energy cosmic ray such as neutrino comes into the core, it interacts with a light quark in the matter and transforms the light quark to a charm quark though weak interaction.
In this situation, QCD Kondo effect and/or magnetically induced QCD Kondo effect can occur inside neutron star and magnetar.

Another situation is relatively low energy heavy ion collisions to be conducted in GSI-FAIR and J-PARC.
There, a quark matter at high density and low temperature would be created.
Hard processes at the initial stage of the collision can produce charm quark pairs, and they induce QCD Kondo effect in the quark matter created.
In this situation, let us focus on the anti-charm quark.
In the hadronization process, the anti-cham quark must be either coupled to the charm quark being charmonium or to light quark being $D$-meson.
If the QCD Kondo effect occurs, the fraction of $D$-meson production should increase.
This will be reflected on the ratio of the charmonium and $D$-meson yields w/ and w/o QCD Kondo effect, and can be a strong signature of the QCD Kondo effect.

At finite temperature, the temperature dependence of resistivity of QCD Kondo effect is analyzed recently in Ref.~\cite{Yasui:2017bey}.
It is shown that the resistivity of QCD Kondo effect increases as the temperature decreases.
In non-perturbative regions, the exact form of the resistivity can be found in Ref.~\cite{Affleck:1992ng}, which depends on the fundamental parameters ($N$, $k$, $R_{\rm{imp}}$) given for finite density QCD Kondo effect and magnetically induced QCD Kondo effect in the present analysis.
The enhancement of the resistivity due to the QCD Kondo effect might be able to be observed in asymmetric heavy ion collisions.
In Ref.~\cite{Hirono:2012rt}, the authors investigate a possibility of observing the conductivity (or equivalently resistivity) of quark-gluon plasma (QGP) in asymmetric heavy ion collisions owing to the imbalance of the charge distributions of the two nuclei.
In this situation, an enhancement of the resistivity could be also a signal of the QCD Kondo effect.

Apart from QCD phenomenology, ``QCD Kondo effect'' would be realized in several situations in condensed matter physics.
From the viewpoint of universality of the fixed point, the finite density QCD Kondo effect is classified into the $\SU(3)$ Kondo universality class, whose realization has been recently proposed in condensed-matter systems, e.g. quantum dot~\cite{Lopez:2013PRB}, ultracold atomic systems~\cite{Nishida:2013PRL,Nishida:2016PRA}.
Another recent proposal of the $\SU(3)$ Kondo system based on the Tomonaga--Luttinger liquid description directly gives rise to the two-channel system~\cite{Hu:2016QBM}, whose universality is exactly the same as the finite density QCD Kondo effect for $N_f = 1$ theory.
In addition, the three-channel Kondo effect caused by ordinary $\SU(2)$ spin interaction is also relevant to our $\SU(3)$ problem, through the level-rank duality between $\widehat{\SU(3)}_2$ and $\widehat{\SU(2)}_3$.
This implies that various aspects of finite density QCD Kondo effect could be examined in experiments of these systems.
It would be helpful for understanding of QCD Kondo effect, and our CFT approach can be directly applied to such systems.

As mentioned in Sec.~\ref{sec:in-medium} and Sec.~\ref{sec:mag-system}, we have observed the symmetry enhancement in the effective (1+1)-dimensional model of QCD Kondo effect.
In particular, we have shown the color symmetry enhancement in the strong magnetic field system, due to the LLL degeneracy.
This allows us to apply the large $N$ analysis in the large $B$ limit.
Let us comment on several arguments peculiar to this limit.
According to the Mermin--Wagner theorem, there is no spontaneous symmetry breaking in a (1+1)-dimensional system~\cite{Mermin:1966fe,Hohenberg:1967zz,Coleman:1973ci}.
However, the large $N$ limit suppresses the long-range fluctuation, so that a second order phase transition becomes possible.
In fact, the large $N$ Kondo model exhibits a phase transition where the Kondo singlet plays a role of the order parameter~\cite{Coleman:1986dva,Coleman:1987PRB}, while it is a cross-over transition in a realistic situation.
We can expect that such a phase transition actually occurs in the strong magnetic field Kondo effect, and it would be relevant to the recent work on the Kondo phase diagram~\cite{Yasui:2016svc}.
In addition to the standard saddle point approximation, in the large $N$ limit, we can also apply another non-perturbative method, called the AdS/CFT correspondence, to the Kondo problem~\cite{Erdmenger:2013dpa,O'Bannon:2015gwa,Erdmenger:2015spo,Erdmenger:2016vud,Erdmenger:2016jjg,Erdmenger:2016msd}.
See also~\cite{Padhi:2017uxc}.
This would be also interesting since the QCD Kondo effect provides a possible realistic application of the AdS/CFT correspondence.

In the QCD Kondo effect, the color exchange interaction between quarks provides a non-Abelian property needed for the appearance of the Kondo dynamics.
Actually, instead of the color exchange interaction, the iso-spin exchange interaction between nucleon and heavy flavor hadrons can play a similar role.
By using the iso-spin exchange interaction, Yasui and Sudoh discuss possibilities of Kondo effect realized in nuclear matter with a heavy flavor hadron as an impurity~\cite{Yasui:2013xr, Yasui:2016ngy,Yasui:2016hlz,Yasui:2016yet,Yasui:2017izi,Suzuki:2017gde}.
See also \cite{Kanazawa:2016ihl} for another realization of Kondo effect.
We expect that the CFT approach can be also applied to the Kondo dynamics in nuclear physics.

In this paper we have only discussed the one-impurity Kondo problem.
However, the interaction between impurity spins induced by the conduction electrons, called the Ruderman--Kittel--Kasuya--Yoshida (RKKY) interaction, would play an important role for heavy electron systems.
The minimal model describes the interaction between the impurities is the two-impurity Kondo model, which is also studied based on the CFT analysis~\cite{Affleck:1991yq,Affleck:1995PRB}.
It would be interesting to consider a similar situation for QCD Kondo systems where light quarks induce an interaction with heavy quarks, in addition to the ordinary color interaction.

\subsection*{Acknowledgements}

We would like to thank T. Hayata, Y. Hidaka, and M. Nitta for useful discussion.
The work of TK and SO was supported in part by MEXT-Supported Program for the Strategic Research Foundation at Private Universities ``Topological Science'' (No.~S1511006).
TK was also supported in part by Keio Gijuku Academic Development Funds, JSPS Grant-in-Aid for Scientific Research (No.~JP17K18090), and JSPS Grant-in-Aid for Scientific Research on Innovative Areas ``Topological Materials Science'' (No.~JP15H05855), and ``Discrete Geometric Analysis for Materials Design'' (No.~JP17H06462).

\newpage
\appendix


\bibliographystyle{ytphys}
\bibliography{qcdkondo}

\end{document}